\documentclass[pre]{revtex4-1}
\usepackage{amsmath,amssymb,latexsym,epsfig,graphics,epsf,bm}
\usepackage{graphics}
\usepackage{float}
\newcommand{\br}{\bold r}

\renewcommand{\tt}{\tilde{t}}

\newcommand{\rhoet}{\rho_{\rm 1}}
\newcommand{\rhoto}{\rho_{\rm 2}}
\newcommand{\rhoa}{\rho_{\rm a}}
\newcommand{\rhob}{\rho_{\rm b}}
\newcommand{\pet}{p_{\rm 1}}
\newcommand{\pto}{p_{\rm 2}}

\newcommand{\rhoiet}{\rho_{\rm\, 1a}}
\newcommand{\rhoito}{\rho_{\rm\, 1b}}

\newcommand{\rhofet}{\rho_{\rm\, 2a}}
\newcommand{\rhofto}{\rho_{\rm\, 2b}}

\newcommand{\Tet}{T_{\rm 1}}
\newcommand{\Tto}{T_{\rm 2}}
\newcommand{\Ta}{T_{\rm a}}
\newcommand{\Tb}{T_{\rm b}}
\newcommand{\Tiet}{T_{\rm\, 1a}}
\newcommand{\Tito}{T_{\rm\, 1b}}
\newcommand{\Tfic}{T_{\rm f}}

\newcommand{\Tfet}{T_{\rm\, 2a}}
\newcommand{\Tfto}{T_{\rm\, 2b}}

\newcommand{\Ua}{U_{\rm a}}
\newcommand{\Ub}{U_{\rm b}}

\newcommand{\bR}{\bold R}

\newcommand{\bet}{\bm{\eta}}
\newcommand{\tbeta}{\tilde{\bm{\eta}}}

\newcommand{\bRa}{{\bf R}_{\rm a}}

\newcommand{\bRb}{{\bf R}_{\rm b}}

\newcommand{\tnab}{\tilde\nabla}
\newcommand{\tbR}{{\tilde{\bold R}}}

\newcommand{\be}{\begin{equation}}
\newcommand{\ee}{\end{equation}}
\newcommand{\fig}[1]{Fig.~\ref{#1}}
\newcommand{\eq}[1]{Eq.~(\ref{#1})}
\newcommand{\Eq}[1]{Equation~(\ref{#1})}
\newcommand{\CVex}{{C_V}^{\rm ex}}
\newcommand{\tCVex}{{\tilde{C}_V^{\rm ex}}}

\newcommand{\Sex}{{S}_{\rm ex}}
\newcommand{\Sexf}{{S}_{\rm ex,\,2}}
\newcommand{\Sexi}{{S}_{\rm ex,\,1}}
\newcommand{\tSex}{\tilde{S}_{\rm ex}}
\newcommand{\Teq}{{T}_{\rm s}}

\begin{document}

\title{Isomorph theory of physical aging}
\date{\today}
\author{Jeppe C. Dyre}
\email{dyre@ruc.dk}
\affiliation{Glass and Time, IMFUFA, Department of Science and Environment, Roskilde University, P.O. Box 260, DK-4000 Roskilde, Denmark}

\begin{abstract}
This paper derives and discusses the configuration-space Langevin equation describing a physically aging R-simple system and the corresponding Smoluchowski equation. Externally controlled thermodynamic variables like temperature, density, pressure enter the description via the single parameter $\Teq/T$ in which $T$ is the bath temperature and $\Teq$ is the ``systemic'' temperature defined at any time $t$ as the thermodynamic equilibrium temperature of the state point with density $\rho(t)$ and potential energy $U(t)$. In equilibrium $\Teq\cong T$ with fluctuations that vanish in the thermodynamic limit. In contrast to Tool's fictive temperature and other effective temperatures in glass science, the systemic temperature is defined for any configuration with a well-defined density, even if it is not in any sense close to equilibrium. Density and systemic temperature define an aging phase diagram in which the aging system traces out a curve. Predictions are discussed for aging following various density-temperature and pressure-temperature jumps from one equilibrium state to another, as well as for a few other scenarios. The proposed theory implies that R-simple glass-forming liquids are characterized by a dynamic Prigogine-Defay ratio of unity. 
\end{abstract}
\maketitle

\section{Introduction}\label{intro}

Aging is the general term used for gradual changes of material properties. In practice aging is often caused by chemical reactions, but in certain cases the ``physical'' aging due to slight adjustments of molecular positions is more important \cite{scherer}. For decades  phenomenological models have been used in industry to predict the physical aging of inorganic glasses and polymers, both during production and in subsequent use \cite{kov63,moy76,str78,scherer,hut95,mck95,mau09a,can13,rut17}. A number of theories of physical aging have been developed in different contexts \cite{too46,nar71,moy76,scherer,cug94,hod95,ols98,cug99,kob00,ado07,cas07,mau09a,par09,kol12,hec15}, but there are still fundamental scientific challenges and limitations to the models used. Given this fact and the significant technological interest in the subject it is surprising that physical aging is not more widely studied in academia. This may be because aging experiments are quite challenging. A setup studying minute changes of material properties, which take place over weeks and months, severely limits the number of experiments that can be carried out. This frustrates the experimentalist when something goes wrong or it turns out, for instance, that a slightly different annealing temperature should have been used and months may have been wasted.

The present paper is motivated by Niss' recent study of physical aging of molecular weight 390 polyisobutylene, a seven unit ``polymer'' \cite{nis17}. Her experiments utilized high-precision dielectric spectroscopy to monitor slight density changes following temperature jumps small enough to be virtually linear, i.e., below 0.2 K, as well as jumps resulting in genuinely nonlinear structural relaxation (max jump: $2$K). The findings were rationalized by assuming that aging states can be mapped onto the equilibrium temperature-density phase diagram. In this diagram Niss proposed the existence of ``isostructural'' lines along which the system jumps after a temperature or pressure change, subsequently relaxing toward equilibrium. 

The existence of isostructural lines in the equilibrium thermodynamic phase diagram is a prediction of the isomorph theory \cite{IV}, so an obvious question that arises is: Can Niss' physical picture be interpreted within this framework? The present paper develops a general isomorph theory of physical aging and shows that it gives rise to a mapping into a phase diagram similar to that envisaged by Niss. The proposed framework is limited to systems obeying the isomorph theory, though, i.e., those with hidden scale invariance. Such so-called R-simple systems \cite{mal13,abr14,fer14,fle14,pra14,buc15,har15,hey15,sch15a,adr16,khr16,mau16,oza16,rom17,shr17} include the solid and liquid phases of most metals, van der Waals bonded systems, and weakly ionic or dipolar systems, whereas systems with strong directional bonds like covalently or hydrogen-bonded glass-forming liquids are not R-simple \cite{dyr14}. In particular, the traditional oxide glasses are not expected to be described by the theory developed below.

After briefly reviewing the isomorph theory in Sec. \ref{overview}, Sec. \ref{basics} derives the general equations describing the physical aging of R-simple systems within a Langevin equation framework, which gives rise to the concept of a systemic temperature. Section \ref{Syst_T} compares the systemic temperature to Tool's fictive temperature from 1946 and introduces an ``aging phase diagram'' defined by density and systemic temperature. Section \ref{jumps} discusses predictions for different scenarios embodying a sudden change of thermodynamic parameters to new, constant values, and Sec. \ref{other} discusses a few other predictions. Finally, Sec. \ref{summary} gives a brief discussion. The Appendix connects to a previous characterization of single-order-parameter systems in terms of their linear-response thermoviscoelastic properties \cite{ell07} by showing that the theory developed here implies unity dynamic Prigogine-Defay ratio.

\section{Isomorph theory}\label{overview}

If the vector of all $N$ particle coordinates is denoted by $\bR\equiv (\br_1, .. , \br_N)$ and the potential-energy function by $U(\bR)$, an R-simple system by definition obeys the following condition for uniform scaling of same-density configurations $\bRa$ and $\bRb$ \cite{sch14}:

\be\label{Rdef}
U(\bRa)< U(\bRb)\Rightarrow U(\lambda\bRa)< U(\lambda\bRb)\,.
\ee
Here $\lambda$ is a scaling parameter. For realistic R-simple system the scale-invariance property \eq{Rdef} is only obeyed for $\lambda$'s not to different from unity and only for the majority of the system's physically relevant configurations; how well the theory applies depends on the state points in question. For most systems obeying \eq{Rdef} to a good approximation this is not obvious from the mathematical expression for the potential energy, which is why the term ``hidden scale invariance'' is sometimes used \cite{dyr14}.

R-simple systems have isomorphs in their thermodynamic phase diagram, which are lines along which the system's structure and dynamics are invariant to a good approximation \cite{dyr14}. Due to its isomorphs an R-simple system has a phase diagram that is basically one-dimensional in regard to many properties. This excludes complex behavior; hence the name ``R-simple'' for systems that have isomorphs (the term ``simple system'' implies a system of particles interacting via pair potentials \cite{ric65,tem68,wca,big75,sti75,ros79a,bau80,barrat,hey05,kir07,mal07,kre09,ing12,dyr16}).

Isomorph theory is based on the use of reduced units. These are macroscopically defined and different from those usually used in reporting results from computer simulations based on the parameters of the interaction potential. At any given thermodynamic state point reduced units are defined from the density-dependent length $l_0$ in which $\rho=N/V$ is the particle number density:
                                                                              
\be\label{l_unit}
l_0\equiv\rho^{-1/3}\,,
\ee  
the thermal energy $e_0$ in which $T$ is the temperature:

\be\label{e_unit}
 e_0\equiv k_BT\,,
\ee
and the time $t_0$ in which $\mu$ is the generalized mobility defined in the Langevin equation \eq{Lang1} below:

\be\label{t_unit}
t_0\equiv l_0^2/(\mu e_0)=\rho^{-2/3}/(\mu k_BT)\,.
\ee
These units are state-point dependent, but experimentally accessible without knowing the system's Hamiltonian (Newtonian dynamics leads to the different time unit $t_0=\rho^{-1/3}\sqrt{m/k_BT}$ where $m$ is the average particle mass \cite{IV,dyr16}).

Recall that the excess entropy $\Sex$ is the entropy minus that of an ideal gas at the same temperature and density \cite{han13} ($\Sex<0$ because a liquid is always more ordered than a gas). The isomorph concept is derived from \eq{Rdef} as follows. We define the microscopic excess entropy function $\Sex(\bR)$ as the thermodynamic excess entropy of the equilibrium state with density $\rho$ and average potential energy $U(\bR)$ \cite{sch14}. Thus two configurations have the same excess entropy if they have same density and potential energy (it is assumed that all physically relevant configurations fill out the volume $V$ with no holes and thus define a unique density). Utilizing the microcanonical expression for the excess entropy $\Sex$, Ref. \onlinecite{sch14} showed from \eq{Rdef} that $\Sex(\bR)$ depends only on a configuration's \textit{reduced} coordinate vector $\tbR\equiv\bR/l_0= \rho^{1/3}\bR$, i.e.,

\be\label{sextil}
\Sex(\bR)\,=\,\Sex(\tbR)\,.
\ee
If $U(\rho,\Sex)$ is the thermodynamic average potential energy regarded as a function of density and excess entropy, the definition of $\Sex(\bR)$ in conjunction with \eq{sextil} leads to

\be\label{Rfundeq}
U(\bR)
\,=\, U(\rho,\Sex(\tbR)) \,.
\ee
\Eq{Rfundeq} implies invariance of the reduced-unit structure and dynamics along the curves of constant $\Sex$ in the thermodynamic phase diagram \cite{sch14}. These curves are termed isomorphs \cite{IV,sch14}.

Although the isomorph theory is exact only for systems with an Euler-homogeneous potential-energy function plus a constant, its predictions have been confirmed in computer simulations of Lennard-Jones type systems \cite{IV,V}, simple molecular models \cite{ing12b}, crystals \cite{alb14}, nano-confined liquids \cite{ing13a}, non-linear shear flows \cite{sep13}, zero-temperature plastic flows of glasses \cite{ler14}, polymer-like flexible molecules \cite{vel14,vel15a}, metals studied by DFT \textit{ab initio} computer simulations \cite{hum15}, plasmas \cite{vel15}, non-viscous liquids \cite{dyr16,bac14a}, and the Lennard-Jones fluid in four dimensions \cite{cos16a}. Moreover, the theory recently provided the basis for quantitative predictions for the thermodynamics of freezing and melting and how various quantities change along the melting line \cite{cos16,ped16}. Experimental confirmations of isomorph-theory predictions have been presented in Refs. \onlinecite{nie08} and \onlinecite{gun11,roe13,xia15,nis17,han18}. In particular, the density-scaling relation obeyed by many glass-forming liquids \cite{alb04,rol05,rol08a,rol12} -- as well as the so-called isochronal superposition property \cite{rol03,nga05,xia15,adr16} -- are both consequences of the theory \cite{IV}.

Consider now two state points $(\rho_1,T_1)$ and $(\rho_2,T_2)$ on the isomorph with excess entropy $\Sex^0$, and suppose $\bR_1$ and $\bR_2$ are configurations of these state points corresponding to different densities but with the same reduced coordinates, i.e., $\tbR_1=\tbR_2$. Since $(\partial U/\partial\Sex)_\rho=T$ \cite{sch14} first-order Taylor expansions of \eq{Rfundeq} lead to

\be\label{firstorder}
\begin{split}
U(\bR_1)\,&=\,U_1+T_1\left({\Sex(\tbR_1)-\Sex^0}\right)\\
U(\bR_2)\,&=\,U_2+{T_2}\left({\Sex(\tbR_2)-\Sex^0}\right)\,.
\end{split}
\ee
Here $U_1$ and $U_2$ are the average potential energies at the two state points (henceforth, whenever we write a quantity without reference to a configuration $\bR$, the equilibrium thermal average of this quantity at the state point in question is implied, e.g., $U=\langle U(\bR)\rangle$, etc). Eliminating $\Sex(\tbR_1)-\Sex^0=\Sex(\tbR_2)-\Sex^0$ one gets

\be\label{fi_app}
\frac{U(\bR_1)-U_1}{T_1}
\,=\,\frac{U(\bR_2)-U_2}{T_2}\,.
\ee
In this approximation the Boltzmann canonical probabilities of configurations from two isomorphic state points, which can be scaled uniformly into one another, are thus proportional \cite{IV}, i.e.,

\be\label{iso_def}
e^{-U(\bR_1)/k_BT_1}
\,=\,C_{12}\,e^{-U(\bR_2)/k_BT_2}\,\,\,{\rm whenever}\,\,\,\rho_1^{1/3}\bR_1=\rho_2^{1/3}\bR_2\,.
\ee
Here $C_{12}$ is a constant that does not depend on the two configurations. The formalism developed below assumes the first-order expansion \eq{firstorder} and, consequently, the invariance of the canonical probabilities of scaled configurations along an isomorph that follows from \eq{iso_def}. 

Important roles are played in the isomorph theory by the potential energy $U$ and the virial $W$. Recall that the latter quantity gives the term added in the ideal-gas equation to reflect particle interactions \cite{han13,tildesley}:

\be\label{virial}
pV
\,=\,Nk_BT+W\,.
\ee
The microscopic virial $W(\bR)$ is defined from the potential-energy change for a uniform scaling of all particle coordinates \cite{tildesley}, i.e., keeping $\tbR$ fixed:

\be\label{virial_def}
W(\bR)\,\equiv\, \left(\frac{\partial U(\bR)}{\partial\ln\rho}\right)_{\tbR}\,.
\ee
\Eq{Rfundeq} implies 

\be\label{virial_ident}
W(\bR)
\,=\,\left.\frac{\partial U(\rho,\Sex)}{\partial\ln\rho}\right|_{\Sex=\Sex(\tbR)}\,.
\ee
Thus the microscopic virial has a form analogous to that of the potential energy (\eq{Rfundeq}): $W(\bR)=W(\rho,\Sex(\tbR))$ in which $W(\rho,\Sex)$ is the thermodynamic average virial as a function of density and excess entropy. Since $\Sex(\tbR)$ is by definition a function of $\rho$ and $U(\bR)$, $W(\bR)$ may also be regarded as a function of $\rho$ and $U(\bR)$. Summarizing one has

\be\label{W_rel}
W(\bR)
\,=\,W(\rho,\Sex(\tbR))
\,=\,W(\rho,U(\bR))\,.
\ee
Via \eq{virial} this implies for the pressure of the configuration $\bR$ of density $\rho$

\be\label{vir2}
p(\bR)
\,=\,\rho\,\Big(k_BT\,+\,W(\rho,\Sex(\tbR))/N\Big)\,.
\ee
The above equations implicitly assume that the configurations are typical for a liquid or a solid, i.e., do not have large holes and fill out space uniformly. In this way any given configuration $\bR$ defines a density $\rho(\bR)$ though we for simplicity just write $\rho$.

\Eq{W_rel} implies perfect correlations between virial and potential energy fluctuations at constant density \cite{sch14}, the property that originally defined an ideal R-simple (strongly correlating) system \cite{ped08,I,II}. Since the isomorph theory is usually only approximate, \eq{W_rel} does not apply rigorously for all configurations and at all densities. This means that realistic R-simple systems are characterized by strong, but not perfect correlations between the virial and potential-energy constant-volume equilibrium fluctuations \cite{I,dyr14}.

When applying the below aging theory to density and pressure jumps in Sec. \ref{jumps} we make use of the fact that compressing an R-simple system from the outside results in a uniform scaling of all particle coordinates. This follows from \eq{Rfundeq} and \eq{W_rel}, which imply that a uniform compression results in a force distribution throughout the sample that is proportional to the original one. An alternative way of proving uniform compression for R-simple systems proceeds via the fact that the reduced forces are functions only of the reduced coordinates \cite{dyr16}. 

The uniform compression requirement is not restrictive. In particular, this requirement does not imply spatial homogeneity of the forces between particles, and the R-simple system in question may very well be characterized by force-chains as found, e.g., in granular media \cite{liu95}. An example of this is a mixture of different particles interacting via inverse-power-law pair potentials. This system rigorously obeys the uniform compression requirement, but may nevertheless have nearest-neighbor forces varying by orders of magnitude, depending on the range of pair-potential parameters.

For more on the isomorph theory and its applications to different fields the reader is referred to the reviews given in Refs. \onlinecite{ped11,ing12,dyr14,dyr16}.

\section{Physical aging of R-simple systems: General formalism}\label{basics}


In experimental studies of aging the temperature $T$ is externally controlled and identified as the phonon ``bath'' temperature measured on a thermometer. This quantity is defined whenever there is thermal equilibrium among the system's fast degrees of freedom. Given this role of the bath temperature it is simplest to describe the microscopic dynamics by a Langevin equation of motion, also known as Brownian dynamics \cite{reichl}. There is evidence from computer simulations that for glass-forming liquids, i.e., liquids with much longer relaxation times than phonon times, Newtonian, Brownian, and $NVU$ dynamics \cite{NVU_I} give virtually the same physics \cite{gle98,NVU_II}. 

In the Langevin equation the mean particle velocity is proportional to the force (``Aristotle's law''). The actual velocity is the mean velocity plus a white noise term, the magnitude of which is determined by the bath temperature. The Langevin equation is \cite{cha43,reichl} 

\be\label{Lang1}
\dot\bR
\,=\,-\mu\nabla U(\bR)+\bet(t)\
\ee
in which $\mu$ is the generalized mobility, i.e., velocity over force, and the noise vector $\bet(t)$ is composed of Gaussian random variables $\bet_i(t)$ obeying

\be\label{stoj}
\langle \bet_i(t)\bet_j(t')\rangle
\,=\,
2\mu\,k_BT\, \delta_{ij}\delta(t-t')\,.
\ee
The corresponding Smoluchowski equation for the probability distribution $P(\bR,t)$ is the generalized diffusion equation \cite{cha43,reichl}

\be\label{smol1}
\frac{\partial P(\bR,t)}{\partial t}
\,= \,\mu\,\nabla\cdot \Big(\big(\nabla U(\bR)\big)P(\bR,t)+k_BT\nabla P(\bR,t)\Big)\,,
\ee
the equilibrium solution of which is the canonical distribution

\be
P_{\rm eq}(\bR)\,\propto\, e^{-U(\bR)/k_BT}\,.
\ee

The above is general. We now restrict to R-simple systems. First, the Langevin equation is made dimensionless using the reduced units of Eqs. (\ref{l_unit}), (\ref{e_unit}), and (\ref{t_unit}). As above, a tilde signals that the quantity in question is reduced and dimensionless, e.g., $\tbR=\rho^{1/3}\bR$. Equation (\ref{Lang1}) is made dimensionless by multiplying by $t_0/l_0$ on each side after which the left-hand side becomes $(t_0/l_0)\dot{\bR}=d\tbR/d\tt\equiv\dot{\tbR}$. For the first term on the right-hand side, since $\nabla=\tilde{\nabla}/l_0$ it follows from \eq{Rfundeq} that $\nabla U(\bR)=(\Teq(\bR)/l_0)\tnab\Sex(\tbR)$ in which $\Teq(\bR)$ is the thermodynamic equilibrium temperature of the system with density $\rho$ and excess entropy $\Sex(\tbR)$:

\be\label{Teq}
\Teq(\bR)
\,\equiv\,\left.\frac{\partial U(\rho,\Sex)}{\partial\Sex}\right|_{\Sex=\Sex(\tbR)}\,.
\ee
$\Teq$ may be regarded as the system's excess entropy temperature. We refer to $\Teq$ the ``systemic temperature'' because it is a global, not locally defined temperature. 

In equilibrium $\Teq\cong T$ with fluctuations that vanish in the thermodynamic limit. The systemic temperature is a function of the density $\rho$ and of $\Sex(\tbR)$. Equivalently, via \eq{Rfundeq} $\Teq$ may be regarded as a function of the density and potential energy $U(\bR)$. Depending on the situation, one or the other representation is most convenient to use. Summarizing, the systemic temperature is given by (compare the analogous identities for the virial \eq{W_rel})

\be\label{Teqs}
\Teq(\bR)
\,=\, \Teq(\rho,\Sex(\tbR)) 
\,=\, \Teq(\rho,U(\bR))\,.
\ee

After the multiplication by $t_0/l_0$ on the right-hand side of \eq{Lang1}, since $\mu t_0/l_0^2=1/k_BT$ by \eq{t_unit} the first term becomes $-(\Teq/T)\tnab\tSex(\tbR)$ in which $\tSex\equiv\Sex/k_B$. The second term becomes $\tbeta(\tt)\equiv(t_0/l_0)\bet(t)$ with autocorrelation given by (recall that $C\delta(C x)=\delta(x)$)

\be\label{stoj2}
\langle \tbeta_i(\tt)\bet_j(\tt')\rangle
\,=\,
\frac{t_0^2}{l_0^2}\,2\mu\,k_BT\delta_{ij}\delta(t-t')
\,=\,
2\,\delta_{ij}\delta(\tt-\tt')\,.
\ee
In conjunction with \eq{stoj2} the reduced Langevin equation is thus

\be\label{Lang2}
\dot{\tbR}\,=\,
-\frac{\Teq(\rho(\tt),\Sex(\tbR))}{T(\tt)}\,\tilde{\nabla}\tSex(\tbR)\,+\,\tbeta(\tilde t)\,.
\ee
\Eq{Lang2} is the Langevin equation for an R-simple system. It applies generally, i.e., in equilibrium as well as during aging. The corresponding Smoluchowski equation is

\be\label{smol2a}
\frac{\partial P(\tbR,\tt)}{\partial\tt}
\,=\,\tilde{\nabla}\cdot\left(\frac{\Teq(\rho(\tt),\Sex(\tbR))}{T(\tt)}\big(\tilde{\nabla}\tSex(\tbR)\big) P(\tbR,\tt)\,+\,\tilde{\nabla}P(\tbR,\tt)\right)\,.
\ee
Note that at any given time the reduced time is defined by reference to the density and temperature at that time. Thus with \eq{t_unit} in mind the definition of $\tt$ may be written

\be\label{tt_def}
d\tt
\,=\,\frac{dt}{t_0(\rho(t),T(t))}\,.
\ee

The systemic temperature is an intensive quantity and consequently its fluctuations are insignificant in the thermodynamic limit. This suggests a mean-field approximation that replaces  $\Teq(\rho(\tt),\Sex(\tbR))$ by its ensemble average, i.e., 

\be\label{smol2}
\frac{\partial P(\tbR,\tt)}{\partial\tt}
\,=\,\tilde{\nabla}\cdot\left(\frac{\Teq(\tt)}{T(\tt)}\big(\tilde{\nabla}\tSex(\tbR)\big) P(\tbR,\tt)\,+\,\tilde{\nabla}P(\tbR,\tt)\right)\,
\ee
in which $\Teq(\tt)\equiv\int\Teq(\rho(\tt),\Sex(\tbR))P(\tbR,\tt)d\tbR$. This is a good approximation in all situations except in and very close to thermal equilibrium. We end this section by turning to this situation in which one must refer to \eq{Lang2} and \eq{smol2a}. 

In equilibrium $\Teq\cong T$ so in the thermodynamic limit \eq{Lang2} apparently reduces to

\be\label{Lang3}
\dot{\tbR}\,=\,
-\tilde{\nabla}\tSex(\tbR)\,+\,\tbeta(\tilde t)\,.
\ee
This cannot be correct, however, because \eq{Lang3} has no reference to the thermodynamic state point. In fact, \eq{Lang3} implies that all excess entropy values are equally likely: the equilibrium probability distribution of \eq{Lang3}'s corresponding Smoluchowski equation is proportional to $\exp(-\tSex(\tbR))$, the density of states is proportional to $\exp(\tSex(\tbR))$ by the definition of entropy, and their product is a constant. Keeping the factor $\Teq/T$ in \eq{Lang2} is thus necessary when studying equilibrium fluctuations. This factor prevents the excess entropy from drifting away by increasing the damping whenever the excess entropy (equivalently: potential energy) is larger than its state-point average corresponding to $\Teq>T$, thus taking more potential energy away from the system than required to balance the noise. Conversely, the damping is decreased whenever the excess entropy goes below its state-point average, resulting in increasing excess entropy and potential energy.

For a large system in equilibrium at constant volume $\Teq$ may be expanded as follows

\be
\Teq
\,=\, T +\left(\frac{\partial T}{\partial\Sex}\right)_\rho (\Sex(\tbR)-\Sex)\,.
\ee
Since $(\partial\Sex/\partial T)_\rho=\CVex/T$ this implies with $\tCVex\equiv\CVex/k_B$ and $\tSex\equiv\Sex/k_B$ that

\be\label{tft}
\frac{\Teq}{T}
\,=\, 1+\frac{\tSex(\tbR)-\tSex}{\tCVex}\,.
\ee
Hence, \eq{Lang2} becomes 

\be\label{Lang4}
\dot{\tbR}\,=\,-
\Big(1+\frac{\tSex(\tbR)-\tSex}{\tCVex}\Big)
\,\tilde{\nabla}\tSex(\tbR)\,+\,\tbeta(\tilde t)\,.
\ee
The corresponding Smoluchowski equation is

\be\label{smol4}
\frac{\partial P(\tbR,\tt)}{\partial\tt}
\,=\,\tilde{\nabla}\cdot\left(\Big(1+\frac{\tSex(\tbR)-\tSex}{\tCVex}\Big)\big(\tilde{\nabla}\tSex(\tbR)\big) P(\tbR,\tt)\,+\,\tilde{\nabla}P(\tbR,\tt)\right)\,.
\ee
There is now a reference to the state point in question via its reduced excess entropy $\tSex$. \Eq{smol4} implies that the equilibrium distribution is given by

\be\label{eql2}
P_{\rm eq}(\tbR)
\,\propto\,\exp\left(-\tSex(\tbR)-\frac{(\tSex(\tbR)-\tSex)^2}{2\,\tCVex}\right)\,.
\ee
Both \eq{smol4} and its equilibrium solution \eq{eql2} are isomorph invariant because $\tCVex$ is isomorph invariant in the first-order approximation leading to \eq{fi_app} \cite{IV}. Since the density of states is proportional to $\exp(\tSex(\tbR))$, \eq{eql2} implies a Gaussian equilibrium probability distribution of the reduced excess entropy with standard deviation $\tCVex$, compare the discussion of entropy fluctuations in Ref. \onlinecite{lan58}.

\section{Aging phase diagram defined from density and systemic temperature}\label{Syst_T}

In his seminal 1946 paper \cite{too46} Tool defined the ``equilibrium or fictive temperature'' of a glass $\Tfic$ as the ``temperature at which the glass would be in equilibrium if heated or cooled very rapidly to it''. It is a non-trivial assumption that such a temperature exists. The appealing physical idea is that structure may be quantified in terms of a temperature that in equilibrium is  identical to the actual temperature. In the simplest case it is assumed that the glass' volume and temperature determine $\Tfic$ \cite{too46,nar71,bra85,varshneya}. Similar  structural ``effective'' temperatures have been discussed in various contexts \cite{cas03,cri03,han04,pow05,cug11,pug17}.

Tool did not state whether the imagined rapid heating or cooling is supposed to take place at constant pressure or constant volume, but given the ambient pressure conditions of most experiments he most likely had the former in mind. As discussed by Niss \cite{nis17}, the two scenarios differ in important respects that are not accounted for by Tool's fictive temperature concept. Niss concluded that ``the classical fictive temperature definition de facto ignores that the equilibrium phase diagram has two dimensions'' \cite{nis17}. \Eq{Teq} resolves this challenge by introducing the systemic temperature $\Teq$, which allows for describing both density and pressure jumps unambiguously (Sec. \ref{jumps}). The price paid is that $\Teq$ is only a useful concept for R-simple systems which excludes, e.g., the technologically important case of oxide glasses. On the other hand, the definition \eq{Teq} does not assume the system is close to equilibrium as is implicitly assumed in the definition of Tool's fictive temperature $\Tfic$ and most other effective temperatures \cite{cri03,pow05,cug11,pug17}. The systemic temperature also differs from $\Tfic$ in other respects. For instance, due to the entropy associated with the phonon degrees of freedom, $\Teq$ varies with the actual temperature even deep into the glassy state.

Before proceeding we briefly reflect on how the systemic temperature may be calculated from experimental or computer simulation data. The systemic temperature is a new concept with no one-to-one relation to previously discussed structural temperatures like, e.g., the configurational or effective temperature \cite{cas03,cri03,han04,pow05,cug11,pug17}. The definition of $\Teq$ via $\Sex$ given in \eq{Teq} is of little use in practice, but fortunately \eq{Teqs} implies that knowledge of density and potential energy is enough to determine $\Teq$. In a computer simulation one may map out the equilibrium average potential energy as a function of density and temperature. Inverting these data determines the equilibrium temperature as a function of density and potential energy, which is the functional dependence that defines $\Teq$. Once this has been established, at any given time during an aging computer simulation $\Teq$ is given from the density and potential energy. In experiments the situation is more challenging because the potential energy is not directly measurable and some model must be used to estimate it as a function of density and temperature (a further challenge is to monitor the density of the system with sufficient accuracy). 

Because the configuration $\bR$ determines both the density and the systemic temperature $\Teq=\Teq(\rho,U(\bR))$, at any given time an aging system defines a point in the ``aging phase diagram'' defined as the $(\rho,\Teq)$ plane. Just as the equilibrium phase diagram, the aging phase diagram has isomorphs defined as curves of constant excess entropy. This follows from the fact that by inversion of \eq{Teqs} the excess entropy is a unique function of density and systemic temperature: $\Sex(\tbR)=\Sex(\rho,\Teq(\bR))$. Substituting this into \eq{Rfundeq} and \eq{W_rel} one concludes that the aging phase diagram likewise has curves of constant potential energy and curves of constant virial. Note that while these are mathematically well-defined curves in the aging phase diagram, their existence does not imply that the potential energy or the virial are constant during aging.

By definition, the aging phase diagram and the equilibrium $(\rho,T)$ phase diagram have the same isomorphs, iso-potential-energy curves, and iso-virial curves, i.e., these curves fall on top of each other if the $(\rho,\Teq)$ and the $(\rho,T)$ coordinate systems are put on top of each other. In this sense the aging phase diagram realizes Niss' idea of mapping the aging system onto the equilibrium phase diagram. In particular, the aging phase diagram has the isostructural lines  conjectured by Niss -- these are the isomorphs. Note, however, that the iso-virial lines in the aging phase diagram are \textit{not} isobars since the kinetic contribution to the pressure depends on the temperature $T$ (\eq{virial}) that is not represented in the diagram. The aging phase diagram would be more complete if $T$ was added as a third dimension, which e.g. has well-defined isobaric surfaces, but we stick here to defining the aging phase diagram as the two-dimensional $(\rho,\Teq)$ diagram because it can be drawn in the plane. 

To summarize, at any given time an R-simple system defines a point in the aging phase diagram, no matter whether the system is far out of equilibrium, is aging and approaching equilibrium smoothly, or is in thermal equilibrium. In the latter case the system's point in the aging phase diagram stays constant and is given by the equilibrium density and temperature ($\Teq=T$). The only requirement that needs to be obeyed for mapping an R-simple system into its aging phase diagram is the above-mentioned assumption that the system has no holes and homogeneously fills out space to define an overall density.

\section{Predictions for jumps from one state to another}\label{jumps}

An ideal aging experiment starts in equilibrium, changes the thermodynamic conditions instantaneously, and keeps these constant while monitoring the full approach to equilibrium \cite{hec15}. Such jumps are easily carried out in computer simulations, but are difficult to realize even approximately in the laboratory. Approximating an ideal aging jump experiment requires that the new temperature is established uniformly throughout the sample on a time scale much shorter than that of any significant relaxation. This is challenging due to the slowness of heat conduction and to the broad relaxation time spectra involved in aging, stretching to much shorter times than the average structural relaxation time. The strategy used in the \textit{Glass and Time} group is to approach ideal aging conditions by working with thin samples (0.05 mm) and using a Peltier element for controlling the temperature. In this way it is possible to obtain excellent temperature equilibration within a few seconds \cite{hec10,hec15,nis17}.

Below we detail the predictions for R-simple systems subjected to ideal aging experiments. Two different cases are discussed, density-temperature controlled jumps and pressure-temperature controlled jumps. The former are simplest because density is an explicit variable in the aging equation, whereas pressure control implies a constraint that determines how the density evolves with time. Because aging is controlled by $\Teq/T$ (\eq{smol2}), the central quantity to keep track of is the systemic temperature. Whenever $\Teq<T$ the aging system increases its potential energy during aging, whenever $\Teq>T$ the potential energy decreases.

In most aging experiments and theories it is assumed that the structure ages much more slowly than the phonon (vibrational) degrees of freedom, which equilibrate on the picosecond time scale. For glass-forming liquids one often identifies structure by the so-called inherent state, the mechanical-equilibrium configuration in the potential-energy landscape reached by steepest descent from the actual configuration \cite{web85}. After a temperature change the phonon degrees of freedom equilibrate rapidly on a time scale at which the system is still inside the ``basin'' defined by the inherent state. This physical picture is realistic, though for a large system at any given time barrier transitions occur somewhere in the sample, making the picture more blurry.

We henceforth assume the above standard time-scale separation in which the structure ages on a much longer time scale than required for equilibrating the phonon degrees of freedom. We discuss the predictions for $\Teq$ for different types of jumps, starting at $t=0$ from a state indexed 1, instantaneously changing the thermodynamic conditions to a state indexed 2. The final, ``annealing'' temperature is thus denoted by $\Tto$. To be specific, if the jump is induced by changing one or two thermodynamic quantities, these are assumed to increase, e.g., $\Tto\ge\Tet$, $\rhoto\ge\rhoet$, $\pto\ge\pet$. The opposite situations of one or more quantities decreasing is treated analogously.

For each jump three time regimes are considered: 1) right after the jump indicated by writing $t=0^+$, 2) after phonon equilibration, i.e., after a few picoseconds, 3) after full thermal equilibration. Regimes 1) and 2) cannot be distinguished in experiments, but are easily distinguished in computer simulations. Regimes 2) and 3), on the other hand, are well separated in good experiments. Note that the general aging isomorph theory does not imply or require the  time-scale separation that follows from the existence of a well-defined phonon equilibration time scale much shorter than the time of molecular rearrangements. Nevertheless, in order to connect to the experimentally most relevant case we will assume time-scale separation. Another thing to be mentioned is that the discussion below ignores thermal fluctuations. Thus when we write, e.g., $T=\Teq$, it is understood that this applies to the extent that deviations go to zero in the thermodynamic limit.

\subsection{Density-temperature jumps}\label{dtjumps}

\begin{figure}
	\begin{center}
		\includegraphics[width=8cm]{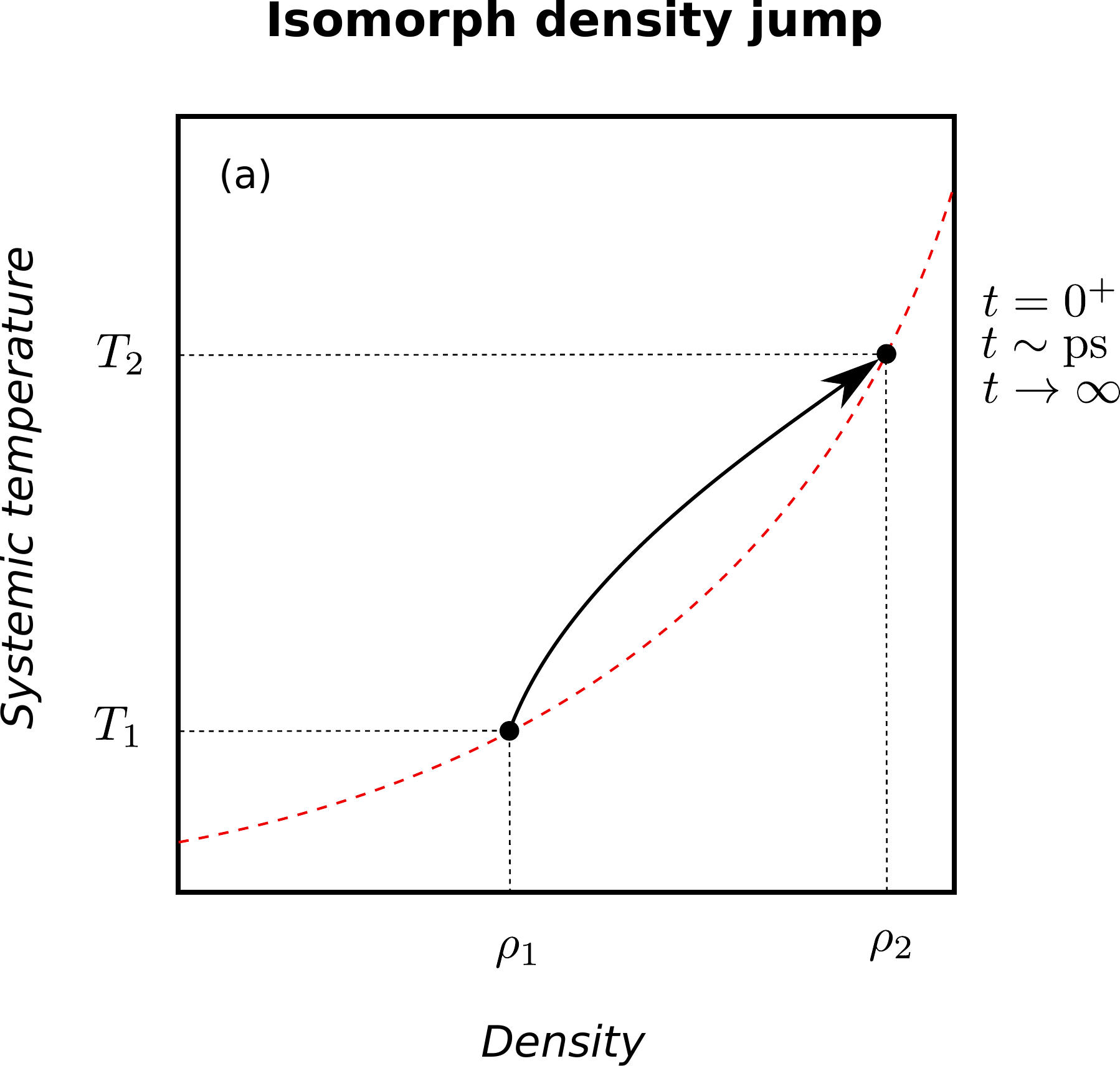} 
		\includegraphics[width=8cm]{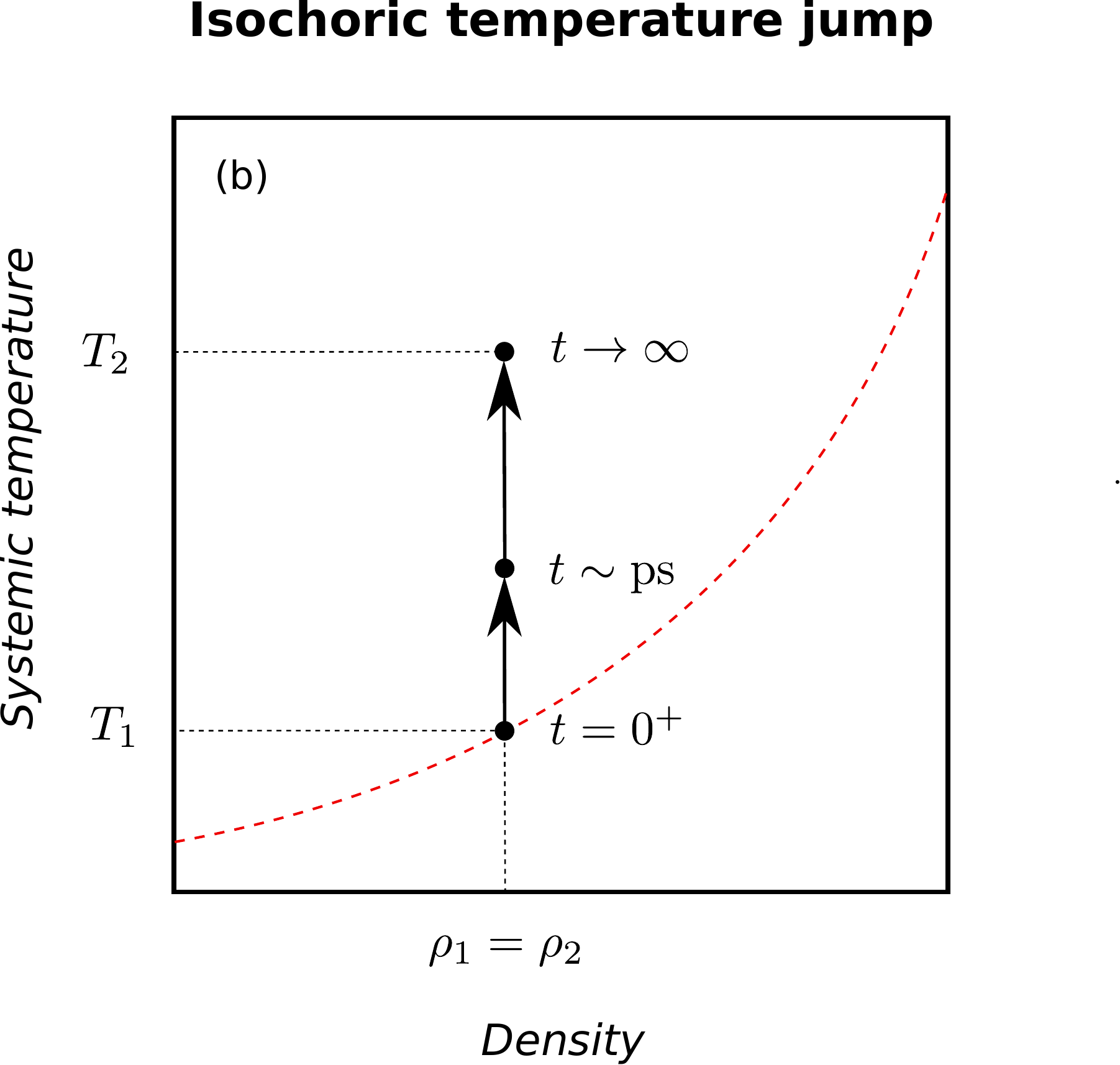} 		
		\includegraphics[width=8cm]{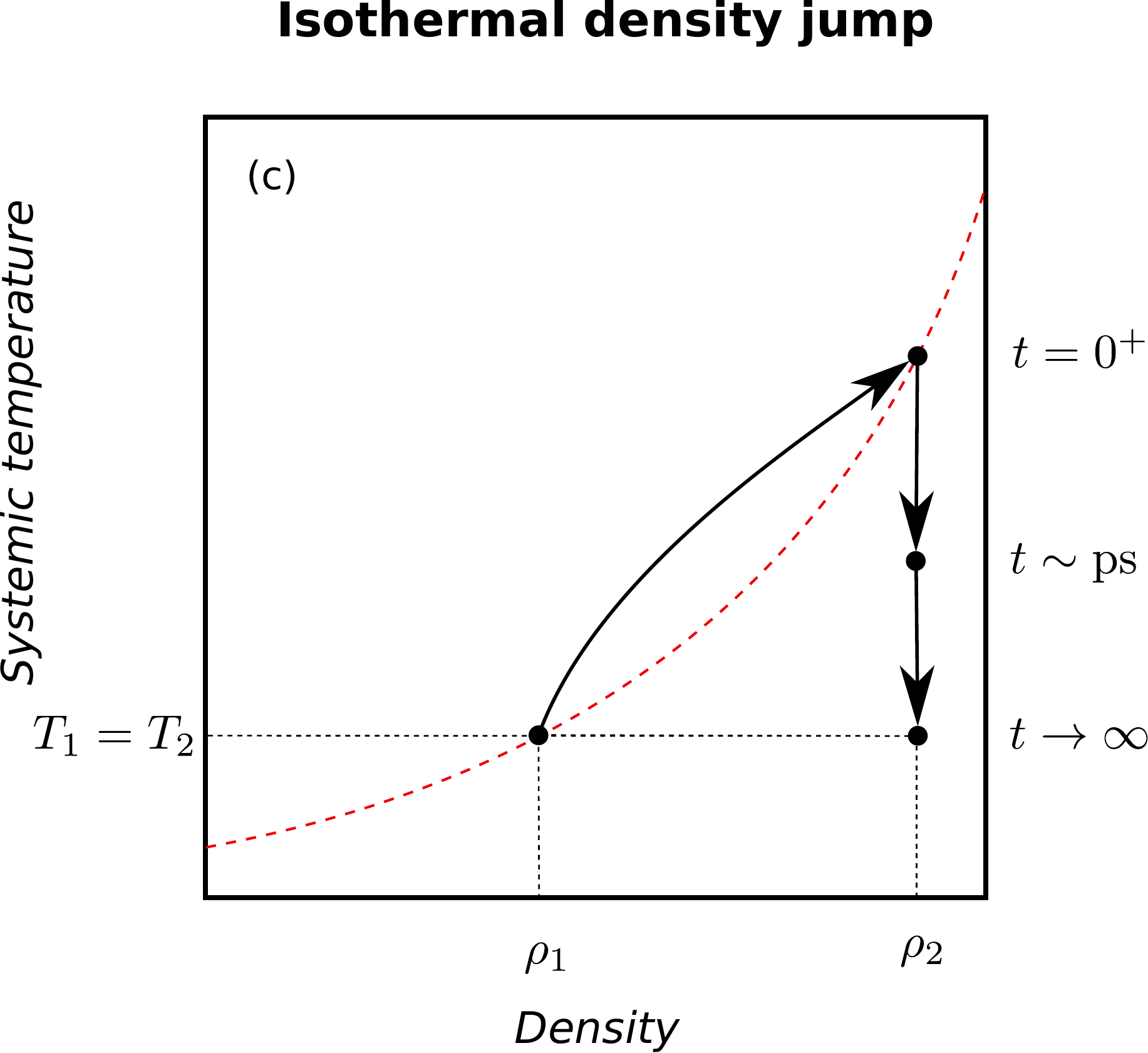} 
		\caption{\label{fig1}\scriptsize  Aging phase diagrams for the three different density-temperature jumps detailed in the text starting in equilibrium at the state point $(\rhoet,\Tet)$ and ending in equilibrium at the state point $(\rhoto,\Tto)$. The figure relates to the typical situation of physical aging in which there is a clear separation of the phonon time scale and the much slower time scale of structural relaxation. The red dashed lines are the isomorphs through the initial state points. States are marked by a black point immediately before and after the jump, after equilibration on the phonon time scale of order picoseconds ($t\sim{\rm ps}$), and when the system is fully equilibrated at which point $\Teq=\Tto$ ($t\rightarrow\infty$). 
		(a) Isomorph density jump. In this case the system is instantaneously in equilibrium at the new density and temperature.
		(b) Isochoric temperature jump. The system does not jump immediately, but gradually  thermalizes by increasing the potential energy (and thus $\Teq$), first on the phonon time scale at which partial equilibration takes place and subsequently as the system equilibrates its structural degrees of freedom.
		(c)	Isothermal density jump. In this case the system is instantaneously compressed to density $\rhoto$ by jumping along the isomorph, after which it subsequently thermalizes.}
	\end{center} 
\end{figure}

This section discusses three different jumps for which density and temperature are the externally controlled thermodynamic variables: an isomorph jump, an isochoric (constant volume) temperature increase, and an isothermal density increase. The jump starts in equilibrium at the state point $(\rhoet,\Tet)$ and ends in equilibrium at $(\rhoto,\Tto)$. It is assumed that the external density control results in a uniform affine transformation of the system, compare the discussion at the end of Sec. \ref{overview}; this implies that right after the jump the system's reduced coordinate $\tbR$ is unchanged. The below predictions are illustrated in \fig{fig1}.

\subsubsection{Isomorph jump}
An isomorph jump takes place between two state points on the same isomorph, i.e., with the same excess entropy:

\be\label{isoj}
\Sex(\rhoet,\Tet)\,=\,\Sex(\rhoto,\Tto)\,.
\ee
In this case, equilibrium is obtained instantaneously at the new state point, no matter how large the equilibrium relaxation time is at the state points in question \cite{IV}. To prove this, note first that right after the jump the density is $\rhoto$ while $\tbR$ and thus $\Sex(\tbR)$ are unchanged, implying that  $\Teq(t=0^+)=\Teq(\rhoto,\Sex(\tbR))$. Before the jump the system is in equilibrium, i.e., $\Sex(\tbR)=\Sex(\rhoet,\Tet)$. From \eq{isoj} we conclude that $\Sex(\tbR)=\Sex(\rhoto,\Tto)$, which means that right after the jump $\Teq(t=0^+)=\Teq(\rho,\Sex(\tbR))=\Teq(\rhoto,\Sex(\rhoto,\Tto))$. According to the definition of the systemic temperature the right-hand side is $\Tto$, implying that

\be
\Teq(t=0^+)=\Tto\,.
\ee
Thus the system is in equilibrium at the new state point $(\rhoto,\Tto)$ right after the jump as far as the systemic temperature is concerned. The equality $\Teq=\Tto$ by itself does not guarantee equilibrium, however. This is ensured by the fact that the equation of motion \eq{Lang2} involves only the \textit{reduced} coordinate, and since $\Teq/T=1$ both before and after the jump, the reduced-unit dynamics is entirely unaffected by the jump. In other words, the equilibrium distribution \eq{eql2} applies before as well as right after the jump. Thereafter, of course, the system stays in equilibrium.

The prediction of instantaneous equilibration for density-temperature isomorph jumps \cite{IV} has been validated in computer simulations of R-simple atomic, molecular, and polymeric model liquids \cite{IV,ing12b,vel14}. Isomorph jumps have also been demonstrated for the Lennard-Jones single crystal studied on the picosecond time scale \cite{alb14}.

\subsubsection{Isochoric temperature jump}

Consider next the situation in which $\rhoto=\rhoet$ and $\Tto>\Tet$. Right after the jump neither the density nor $\Sex(\tbR)$ has changed. Before the jump $\Teq=\Tet$, and since $\Teq$ is a function of density and excess entropy (\eq{Teqs}), we conclude that $\Teq(t=0^+)=\Tet$.

The fact that $\Teq<\Tto$ right after the jump implies that the system on average increases its potential energy when it equilibrates on the phonon time scale. This leads to a stabilization of $\Teq$ on some value obeying $\Tet<\Teq<\Tto$. After this, on the longer time scale of structural equilibration, the system further increases its potential energy until equilibrium has been reached at which $\Teq=\Tto$.

\subsubsection{Isothermal density jump}

In this case $\Tto=\Tet$ and $\rhoto>\rhoet$. Right after the jump $\Sex$ is unchanged, implying that $\Sex(\rhoto,\Teq(t=0^+))=\Sex(\rhoet,\Tet)$. This means that at $t=0$ the system jumps along an isomorph in the aging phase diagram (as shown in Ref. \onlinecite{gna10} this fact may be used to rationalize the long-standing mystery that the effective temperature of a glass in computer simulations depends only on the final density jumped to \cite{dil00}). Because $(\partial T/\partial\rho)_{\Sex}>0$, $\Teq$ jumps at $t=0$ to a larger value: $\Teq>\Tto=\Tet$. When the phonon degrees of freedom subsequently equilibrate, the potential energy decreases. This lowers $\Teq$, initially not to the equilibrium value $\Tto$ that is reached only much later when the structural degrees of freedom equilibrate.

\subsection{Pressure-temperature jumps}\label{ptjumps}

\begin{figure}
	\begin{center}
		\includegraphics[width=8cm]{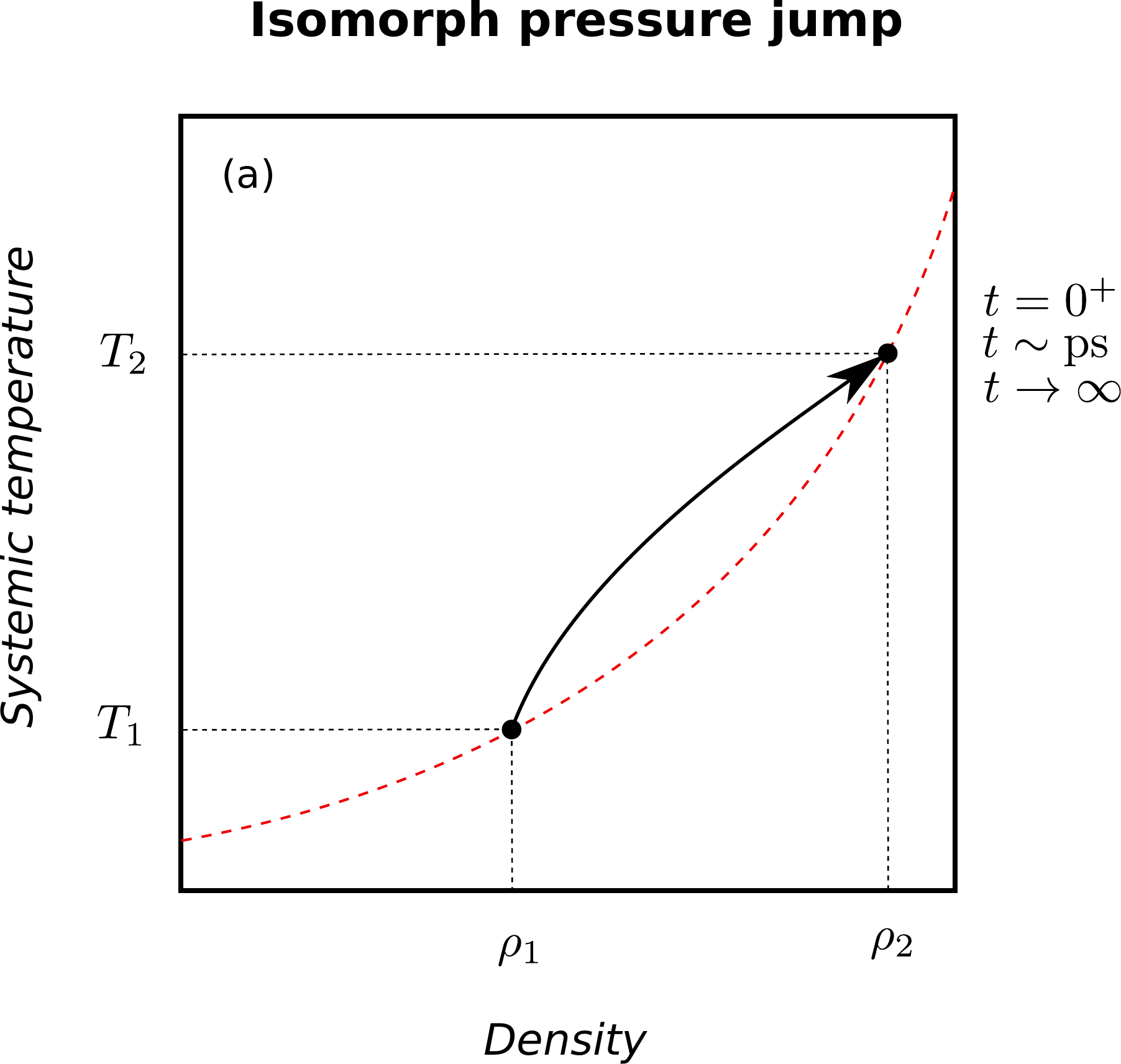} 
		\includegraphics[width=8cm]{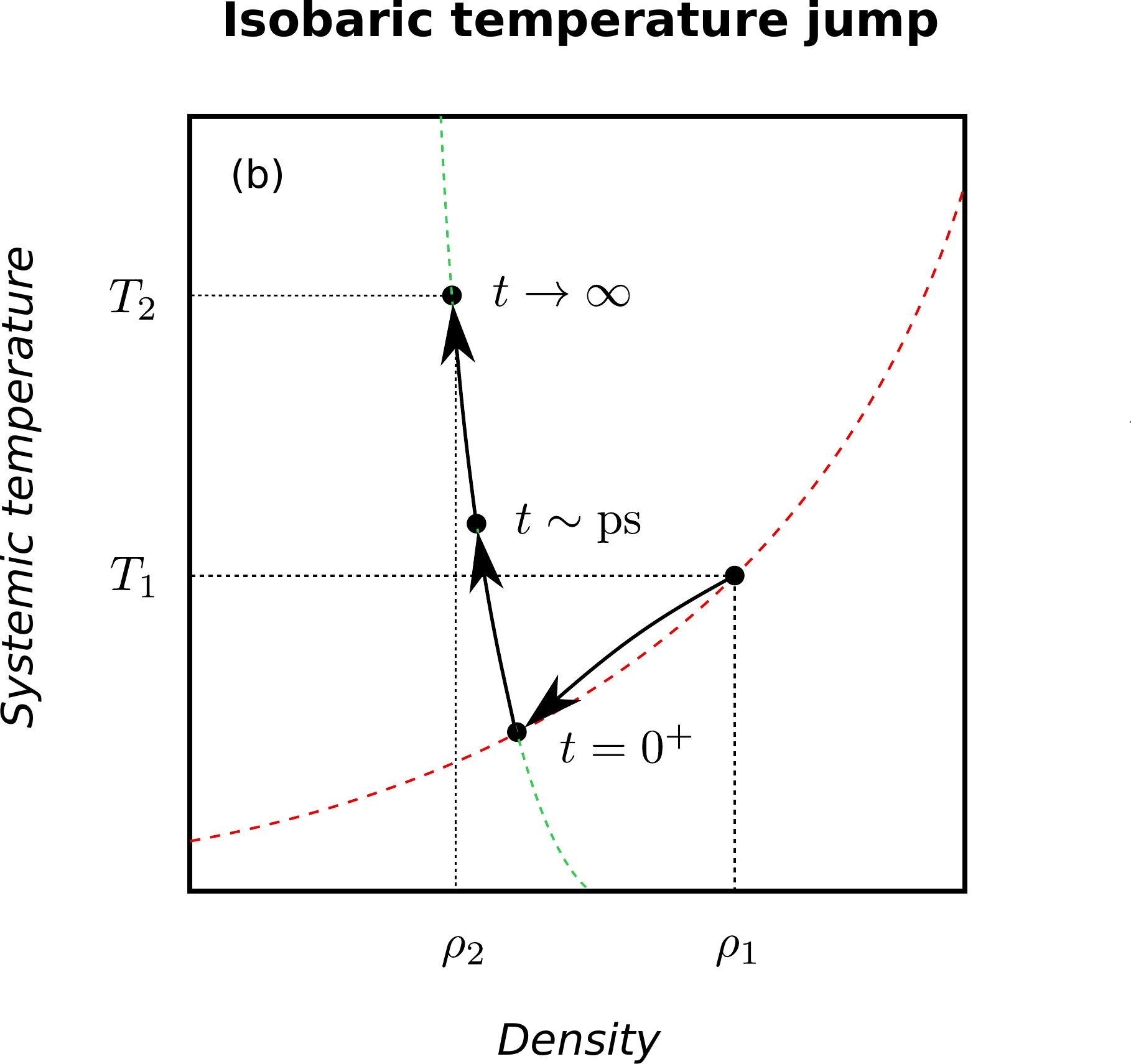} 		
		\includegraphics[width=8cm]{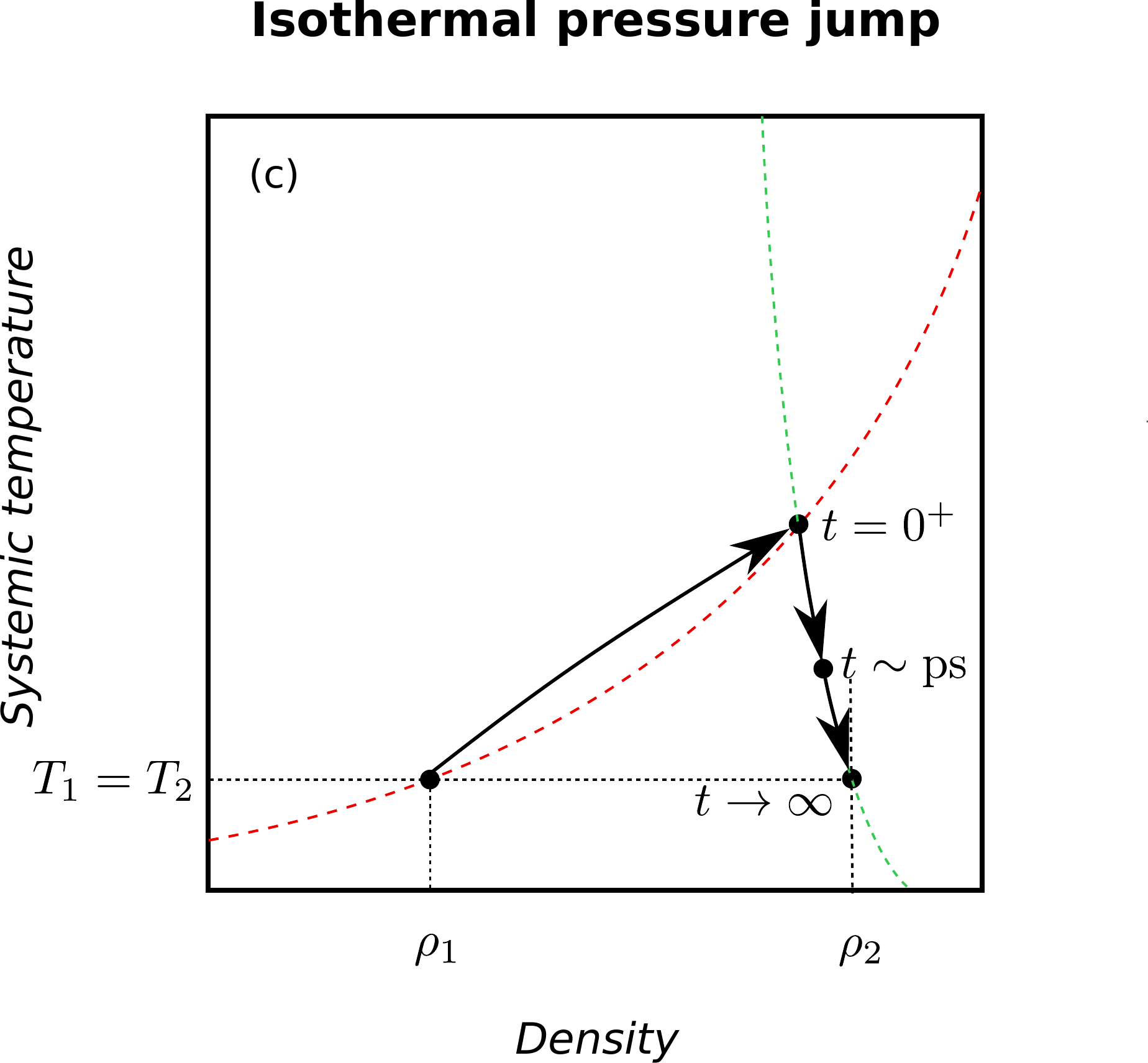} 
		\caption{\label{fig2}\scriptsize Aging phase diagrams for the three different pressure-temperature jumps detailed in the text starting in equilibrium at the state point $(\pet,\Tet)$ and ending in equilibrium at the state point $(\pto,\Tto)$. The corresponding equilibrium densities are denoted by $\rhoet$ and $\rhoto$. The figure relates to the typical situation of physical aging in which there is a clear separation of the phonon time scale and the much slower time scale of structural relaxation. The red dashed lines are the isomorphs through the initial state points, the green dashed lines are isobars defined by $\pto$ and $\Tto$ (\eq{isobar}). States are marked by a black point immediately before and after the jump, after equilibration on the phonon time scale of order picoseconds ($t\sim{\rm ps}$), and when the system is fully equilibrated at which point $\Teq=\Tto$ ($t\rightarrow\infty$). 
			(a) Isomorph pressure jump. In this case the system is instantaneously in equilibrium at the new pressure and temperature, just as for the isomorph density jump.
			(b) Isobaric temperature jump. The system jumps immediately along the isomorph through the initial state point leading to an initial \textit{decrease} of the systemic temperature. After that the system thermalizes by increasing the potential energy and thus $\Teq$ by moving along the isobar defined by $\pto$ and $\Tto$. This happens first on the phonon time scale at which partial equilibration takes place and subsequently as the system equilibrates the structural degrees of freedom.
			(c)	Isothermal pressure jump. In this case there is also an instantaneous isomorph jump, followed by thermalization at constant pressure.
		}
	\end{center} 
\end{figure}

Consider next the common experimental situation in which pressure and temperature are externally controlled. Recall that in terms of the virial $W(\rho,\Sex(\tbR))$ the pressure is given by \eq{vir2}, which involves also the density and temperature. A pressure-temperature jump starts from equilibrium at state point $(\pet,\Tet)$ and ends at equilibrium in $(\pto,\Tto)$. It is assumed that external pressure changes result in affine transformations of the sample (compare the discussion at the end of Sec. \ref{overview}), i.e., that right after the jump the system's reduced coordinate $\tbR$ and thus its excess entropy are unchanged. The below predictions are illustrated in \fig{fig2}.

\subsubsection{Isomorph jump}

A pressure-temperature jump between isomorphic states, i.e., states characterized by $\Sex(\pet,\Tet)\,=\,\Sex(\pto,\Tto)$, leads to instantaneous equilibration just as for a density-temperature isomorph jump. To see this note first that if $\rhoto$ is the equilibrium density of the state point $(\pto,\Tto)$, one has by \eq{vir2}

\be\label{enside}
\pto
\,=\,\rhoto\,\big(k_B\Tto\,+\,W(\rhoto,\Sex(\pto,\Tto))/N\big)\,.
\ee
Since $\Sex(\tbR)=\Sex(\pet,\Tet)$ does not change at $t=0$, the density right after the jump is determined by

\be\label{anside}
\pto
\,=\,\rho(t=0^+)\,\big(k_B\Tto\,+\,W(\rho(t=0^+),\Sex(\pet,\Tet))/N\big)\,.
\ee
For given pressure, temperature, and excess entropy \eq{vir2} determines the density. Comparing \eq{enside} and \eq{anside}, because $\Sex(\pet,\Tet)\,=\,\Sex(\pto,\Tto)$ we conclude that

\be
\rho(t=0^+)\,=\,\rhoto\,.
\ee
This means that after applying the external pressure $\pto$, the system immediately jumps to the equilibrium density at the state point $(\pto,\Tto)$. In effect, the system performs a density-temperature isomorph jump, leading as we have already seen to instantaneous equilibration.

\subsubsection{Isobaric temperature jump}

Consider next the situation in which $\pto=\pet$ and $\Tto>\Tet$. The density jumps to $\rho(t=0^+)$ determined by \eq{vir2},

\be\label{penside}
\pto
\,=\,\rho(t=0^+)\,\Big(k_B\Tto\,+\,W(\rho(t=0^+),\Sex(\tbR))/N\Big)\,.
\ee
Right before the jump one has

\be\label{panside}
\pet
\,=\,\rhoet\,\big(k_B\Tet\,+\,W(\rhoet,\Sex(\tbR))/N\big)\,.
\ee
The temperature increase is compensated by a density decrease, i.e., to keep the pressure constant there is an instantaneous thermal expansion, $\rho(t=0^+)<\rhoet$. 

Since $\Sex$ does not change at $t=0$, the density decrease translates via \eq{Teqs} to a decrease in $\Teq$, i.e., $\Teq(t=0^+)<\Tet$. In effect, the system at $t=0$ performs an isomorph jump taking it to a state of lower density and lower systemic temperature. This initial decrease of the systemic temperature upon isobaric heating may appear counterintuitive, but we note that it is consistent with Niss' discussion of her proposed mapping into the equilibrium phase diagram \cite{nis17}.

The subsequent equilibration takes place along the isobaric curve defined in the $(\rho,\Teq)$ aging phase diagram by $\pto$ and $\Tto$ in which $W(\rho,\Sex)$ is the equilibrium virial function, compare \eq{vir2}: 

\be\label{isobar}
p_2
\,=\,\rho\,\Big(k_BT_2\,+\,W(\rho,\Sex(\rho,\Teq))/N\Big)\,.
\ee
Since $\Teq(t=0^+)<\Tet<\Tto$ one has $\Teq(t=0^+)/\Tto<1$ meaning that the system increases its potential energy when the phonon degrees of freedom equilibrate, i.e., $\Teq$ increases and stabilizes on some value lower than $\Tto$. Finally, as $t\rightarrow\infty$ the system approaches equilibrium and  $\Teq\rightarrow\Tto$. Both equilibration processes take place along the isobaric curve \eq{isobar} defined by $\pto$ and $\Tto$.

\subsubsection{Isothermal pressure jump}

In this case $\Tto=\Tet$ and $\pto>\pet$. Again, because $\Sex$ is continuous at $t=0$, the initial jump takes place along an isomorph. According to \eq{vir2}, in order to increase the pressure the density must jump to a larger value at $t=0$. Since density increases and $\Sex$ is unchanged, $\Teq$ increases, i.e., $\Teq(t=0^+)>\Tto=\Tet$. 

The fact that $\Teq(t=0^+)/\Tto>1$ implies a subsequent decrease in the potential energy, first when the phonon degrees of freedom equilibrate and subsequently when the structural degrees of freedom equilibrate. Both equilibration processes take place along the isobaric curve \eq{isobar} defined by $\pto$ and $\Tto$.

\section{Other predictions}\label{other}

Consider jumps from two state points $(\rhoiet,\Tiet)$ and $(\rhoito,\Tito)$ on an isomorph with excess entropy $\Sexi$ to state points $(\rhofet,\Tfet)$ and $(\rhofto,\Tfto)$, respectively, on a different isomorph with excess entropy $\Sexf$. We now show that these jumps are described by the same equation of motion \eq{Lang2}, i.e., that in this equation the factor $\Teq/T$ is the same at any given reduced time $\tt$ defined from the final state point's density and temperature (\eq{tt_def}). Note first that for any $\Sex$ one has

\be\label{ann_cond}
\frac{\Teq(\rhofet,\Sex)}{\Tfet}
\,=\,
\frac{\Teq(\rhofto,\Sex)}{\Tfto}\,.
\ee
To show this recall that $\CVex=(\partial\Sex/\partial\ln T)_\rho$.  Since $\CVex$ is isomorph invariant in the first-order approximation assumed throughout this paper (Sec. \ref{overview}), $\CVex$ is a function only of $\Sex$ \cite{ing12a}. This implies that $(\partial\ln T/\partial\Sex)_\rho=1/\CVex(\Sex)$. By integration from $\Sexf$ to the arbitrary value $\Sex$ at densities $\rhofet$ and $\rhofto$, respectively, one obtains for the equilibrium temperature function $T(\rho,\Sex)$ and thus for $\Teq(\rho,\Sex)$ 

\be
\ln\Teq(\rhofet,\Sex)-\ln\Teq(\rhofet,\Sexf)
\,=\,\ln\Teq(\rhofto,\Sex)-\ln\Teq(\rhofto,\Sexf)\,.
\ee
Since $\Teq(\rhofet,\Sexf)=\Tfet$ and $\Teq(\rhofto,\Sexf)=\Tfto$, \eq{ann_cond} follows. Substituting into this equation $\Sex=\Sex(\tbR(\tt))$ one concludes that the two jump scenarios, since they start from the same excess entropy, are described by the same equation of motion \eq{Lang2}. Thus the two scenarios age identically as functions of the reduced time $\tt$ \cite{IV}. 

Since a continuous function of the control parameters may be regarded as composed of many small sudden steps, the above generalizes to continuous thermodynamic control-parameter variations.  Suppose that starting from equilibrium at some state point $(\rho,T)$ the system is subjected to two different thermal histories, $(\rhoa(t),\Ta(t))$ and $(\rhob(t),\Tb(t))$. Identical aging behavior is then predicted if and only if one has at all reduced times $\tt$

\be\label{ag_cond1}
\frac{\Teq(\rhoa(\tt),\Ua(\tt))}{\Ta(\tt)}
\,=\,\frac{\Teq(\rhob(\tt),\Ub(\tt))}{\Tb(\tt)}\,.
\ee
It is understood that the reduced units at any given time $t$ are defined by reference to the density and temperature at that time, compare \eq{tt_def}.

The above applies also for constant pressure situations. For instance, if \eq{ag_cond1} is obeyed for experiments cooling a liquid through the glass transition at different pressures, the resulting glasses are predicted to be identical if taken to ambient pressure. Thus no so-called pressure densification \cite{cas17} is predicted for glasses produced by subjecting R-simple glass-forming liquids to cooling histories characterized by \eq{ag_cond1} \cite{cas17}.

\section{Concluding remarks}\label{summary}

This paper has derived the Langevin equation describing physical aging of R-simple systems \eq{Lang2} and its corresponding Smoluchowski equation \eq{smol2}. The external thermodynamic control parameters enter the description via the single number $\Teq/T$ and the formalism confirms the conjecture from 2007 that R-simple systems are single-parameter systems \cite{ell07,ped08a} (Appendix). 

Any R-simple system, in equilibrium as well as out of equilibrium, defines a point in the $(\rho,\Teq)$ aging phase diagram. This phase diagram is close in spirit to that suggested by Niss, who proposed that an aging system may be mapped onto the equilibrium density-temperature phase diagram \cite{nis17}. Niss argued that isostructural lines exist in this phase diagram; these correspond to the isomorphs of the aging phase diagram. A difference is that Niss' phase diagram is the equilibrium phase diagram and consequently has well-defined isobars, whereas the aging phase diagram's isobars depend on the annealing temperature and pressure. Instead, the aging phase diagram has well-defined isovirial lines. A more complete description of aging would be arrived at by mapping the system into a three-dimensional phase diagram with temperature as the third dimension.

For density-temperature jump experiments the aging behavior of R-simple systems depends only on the starting and ending isomorph. Since isomorphs in experiments may be identified as the isochrones (lines of constant relaxation time), this prediction suggests experiments on aging van der Waals molecular glasses or metallic glasses testing the predicted equivalence of different jumps between two isochrones. The isomorph theory is not expected to work for covalently bonded glasses, but it might be worthwhile for comparison to perform similar experiments on such systems. 

Questions for future work include how the Narayanaswamy material time concept \cite{nar71,scherer,hec15} fits into the formalism. Consider for instance the potential-energy clock model \cite{ado07,ado09} according to which the material-time clock rate is controlled by the potential energy. This idea fits nicely into the systemic-temperature concept because the potential-energy clock model implies that the clock rate is a function of the state point in the aging phase diagram, as also predicted by Niss \cite{nis17}. Another open, related question is how the present approach fits into models predicting the time dependence of the viscosity of an aging glass, an important experimental observable in many aging studies \cite{scherer,rut17}.

We finally emphasize that the isomorph theory physical aging is a single-phase theory that ignores the fact that most glass-forming liquids are supercooled, i.e., of higher free energy than the crystalline phase. The statistical mechanics behind the formalism thus ignores the existence of the large part of configuration phase space corresponding to states that contain small or large crystals. This leads to a consistent description, but one may ask: what if local crystal-type fluctuations occur in the supercooled liquid and are important for the physics \cite{yan17}, e.g., for controlling the viscosity? In this case, assuming again the above Langevin equation for the dynamics, there is a large range of parameters for which the systemic temperature is nothing but the melting temperature. This introduces a constant driving force in \eq{smol2} aiming to take the system to lower potential energy by driving it towards crystallization.

\acknowledgments{The author thanks Edan Lerner, Kristine Niss, Lorenzo Costigliola, Nick Bailey, Thomas Schr{\o}der, and Tina Hecksher for inspiration, useful discussions, and comments improving the manuscript. -- This work was supported by the VILLUM Foundation's \textit{Matter} grant 16515.}

\section*{Appendix: Single-order-parameter description and unity dynamic Prigogine-Defay ratio}\label{Niels}

The purpose of this Appendix is to link this paper's reduced-unit Langevin description of the dynamics of R-simple systems to a paper from 2007 \cite{ell07} that proposed a single-frequency criterion for testing whether a glass-forming liquid is described by a single order parameter. 

In the classical definition, a single-order-parameter model of a glass-forming liquid implies exponentially decaying time-autocorrelation functions \cite{dav53}, corresponding to Debye frequency-dependent linear response functions that are rarely observed. Reference \onlinecite{ell07} introduced a more general single-order-parameter concept that allows for non-exponential time-autocorrelation functions. The experimental criterion for a liquid to be described to a good approximation in that framework is that the dynamic Prigogine-Defay ratio is close to unity \cite{ell07}. This is equivalent to the system having strong virial potential-energy correlations, i.e., being R-simple \cite{ped08a,bai08,II}. We proceed to show that the property of a dynamic Prigogine-Defay ratio equal to unity follows from \eq{smol2}.

Consider a system subjected to small periodic temperature and volume perturbations with complex magnitude $\delta T(\omega)$ and $\delta V(\omega)$ around a state of thermal equilibrium (we employ the standard notation of writing, e.g., $T(t)=T_0+\delta T(\omega)\exp(i\omega t)$ in which the real part is implied). Different quantities have been proposed for the single parameter controlling the physics of viscous liquids -- the density, the configurational entropy, the instantaneous shear modulus, etc -- but interestingly the single-parameter assumption may be investigated without knowing the actual nature of the parameter \cite{ell07,hec15}.

In the present context a single-order-parameter system is defined as a system that has some variable $\varepsilon$ with the property that the amplitudes of the periodic entropy and pressure responses induced by small temperature and volume periodic perturbations are given \cite{ell07} by

\be\label{SOP}
\begin{split}
	\delta S(\omega) \,&=\, \gamma_S \delta\varepsilon(\omega)+J_{ST}^\infty \delta T(\omega)+J_{SV}^\infty \delta V(\omega)\\
	\delta p(\omega)\,&=\, \gamma_p \delta\varepsilon(\omega)+J_{pT}^\infty \delta T(\omega)-J_{pV}^\infty \delta V(\omega)\,.
\end{split}
\ee
Here $\gamma_S$ and $\gamma_p$ are real constants and the ``instantaneous'' high-frequency, in-phase responses are characterized by the two-by-two real compliance matrix $J^\infty$ for which Onsager reciprocity implies $J_{SV}^\infty=J_{pT}^\infty$. Relaxation processes are contained in the $\delta\varepsilon(\omega)$ terms. As shown in Ref. \onlinecite{ell07}, \eq{SOP} implies unity dynamic Prigogine-Defay ratio at all frequencies, i.e.,  $-c_V''(\omega)K_T''(\omega)/[T_0(\beta_V''(\omega))^2]=1$ in which $''$ marks the imaginary parts of the following three frequency-dependent thermodynamic linear-response functions: the isochoric heat capacity per unit volume $c_V(\omega)$, the isothermal bulk modulus $K_T(\omega)$, and the isochoric pressure coefficient $\beta_V(\omega)$. 

Assuming the system is R-simple and switching from volume to density, the response of an arbitrary quantity $A$ to externally imposed small temperature and density variations, $T(t)=T_0+\delta T(\omega)\exp(i\omega t)$ and $\rho(t)=\rho_0+\delta \rho(\omega)\exp(i\omega t)$, is now calculated. In general, $A$ depends on the configuration $\bR$ and on the system's thermodynamic state point, i.e., one can write  $A=A(T,\rho,\tbR)$. The entropy as defined here is also of this form since the ideal gas entropy term is a function of temperature and density and $\Sex$ is a function of $\tbR$; the pressure likewise has this structure, compare \eq{vir2}.

If the solution to the Smoluchowski equation \eq{smol2} is denoted by $P(\tbR,t)$, at time $t$ the average of $A$ is given by

\be\label{Aav}
\langle A(t)\rangle
\,=\,\int A(T(t),\rho(t),\tbR)\,P(\tbR,t)\,d\tbR\,.
\ee
The steady-state probability distribution has the form $P(\tbR,t)=P_{\rm eq}(\tbR)+\delta P(\tbR,\omega)\exp(i\omega t)$ in which $P_{\rm eq}(\tbR)$ is the equilibrium probability distribution at the state point $(\rho_0,T_0)$. According to \eq{Aav} to first order the response is given by $\langle A(t)\rangle=\langle A\rangle_{\rm eq}+\delta A(\omega)\exp(i\omega t)$ in which

\be\label{Aresp}
\begin{split}
	\delta A(\omega)
	\,=&\,\int \left[\frac{\partial A}{\partial T}\left(T_0,\rho_0,\tbR\right)\delta T(\omega)
	+ \frac{\partial A}{\partial \rho}\left(T_0,\rho_0,\tbR\right)\delta \rho(\omega)
	\right]P_{\rm eq}(\tbR)\,d\tbR
	\,\\+&\, \int A(T_0,\rho_0,\tbR)\,\delta P(\tbR,\omega)\,d\tbR\,.
\end{split}
\ee
The first integral gives the in-phase $J^\infty$ terms of \eq{SOP}. The non-trivial frequency dependence comes from the second integral. Focusing on this, note that since \eq{smol2} is controlled by $Q\equiv\Teq/T=1+\delta Q$, the steady-state periodic term of the probability amplitude has the following structure: $\delta P(\tbR,\omega)=\Phi(\tbR,\omega)\delta Q(\omega)$. Expanding the virial in \eq{vir2} to first order one gets $p(T_0,\rho_0,\bR)={\rm Const.}+\Lambda (\Sex(\tbR)-\Sex)$. Substituting this into the second integral of \eq{Aresp}, since $\int\delta P(\tbR,\omega)\,d\tbR=0$ one finds the first terms on the right-hand sides of \eq{SOP} with 

\be
\delta\varepsilon(\omega)
\,\propto\,\delta Q(\omega)\int\Sex(\tbR)\, \Phi(\tbR,\omega)\,d\tbR\,
\ee
and $\gamma_p/\gamma_S=\Lambda$. 

How can the use of the mean-field approximation for calculating the linear response close to equilibrium be justified, given that this approximation breaks down in equilibrium (Sec. \ref{basics})? For small but finite perturbations the induced systemic temperature variations are small but finite. This means that for a sufficiently large system the systemic temperature variations are much larger than the equilibrium systemic temperature fluctuations. In other words, the thermodynamic limit is taken before letting the perturbation magnitude go to zero in order to calculate the linear response.

\end{document}